\documentclass[a4paper,11pt]{article}
\usepackage{pos}

\usepackage{caption}
\usepackage{subcaption}

\usepackage{amsmath}

\usepackage{lineno}

\usepackage{setspace}

\title{IceCube Search for High-Energy Neutrinos from Ultra-Luminous Infrared Galaxies}
 \ShortTitle{IceCube Search for Neutrinos from ULIRGs}

\author{The IceCube Collaboration \\{\normalsize \normalfont(a complete list of authors can be found at the end of the proceedings)}}

\emailAdd{pablo.correa@icecube.wisc.edu}

\abstract{With infrared luminosities $L_{\mathrm{IR}} \geq 10^{12} L_{\odot}$, Ultra-Luminous Infrared Galaxies (ULIRGs) are the most luminous objects in the infrared sky. They are predominantly powered by starburst regions with star-formation rates $\gtrsim 100~ M_{\odot}~ \mathrm{yr^{-1}}$. ULIRGs can also host an active galactic nucleus (AGN). Both the starburst and AGN environments contain plausible hadronic accelerators, making ULIRGs candidate neutrino sources. We present the results of an IceCube stacking analysis searching for high-energy neutrinos from a representative sample of 75 ULIRGs with redshift $z \leq 0.13$. While no significant excess of ULIRG neutrinos is found in 7.5 years of IceCube data, upper limits are reported on the neutrino flux from these 75 ULIRGs as well as an extrapolation for the full ULIRG source population. In addition, constraints are provided on models predicting neutrino emission from ULIRGs.

\vspace{4mm}
{\bfseries Corresponding authors:}
Pablo Correa$^{1*}$, Krijn D.~de Vries$^{1}$, Nick van Eijndhoven$^{1}$\\
{$^{1}$ \itshape Vrije Universiteit Brussel, Pleinlaan 2, BE-1050 Brussels, Belgium}\\[4mm]
$^*$ Presenter

\FullConference{37$^{\rm{th}}$ International Cosmic Ray Conference (ICRC 2021)\\
		July 12th -- 23rd, 2021\\
		Online -- Berlin, Germany}

}

\begin{document}
\maketitle

\section{Introduction}
The existence of a diffuse high-energy astrophysical neutrino flux was discovered by the IceCube collaboration in 2013 \cite{IceCube_discovery_science}. However, the origin of this diffuse flux remains largely unknown. Current limits imply that if a single source population is responsible for the diffuse neutrino observations, it should consist of relatively dim but numerous sources \cite{Murase_2016}. Furthermore, the contribution of blazars observed by the \textit{Fermi} Large Area Telescope (LAT) \cite{Fermi_LAT} to the diffuse neutrino flux is limited \cite{IceCube_2LAC_constraints}. Since neutrinos and pionic gamma rays are expected to be produced in the same hadronic interactions, the non-blazar component of the extragalactic gamma-ray background (EGB) above 50 GeV \cite{Fermi_EGB} provides a second constraint to the sources responsible for the diffuse neutrino flux \cite{Bechtol_2017}. In particular, the non-blazar EGB bound hints towards gamma-ray opaque neutrino sources \cite{Murase_2016_hidden_sources}.

Ultra-Luminous Infrared Galaxies (ULIRGs) are the most infrared-luminous objects on the sky, with total infrared (IR) luminosities $L_{\mathrm{IR}} \geq 10^{12} L_{\odot}$ between 8--1000 $\mu$m (see \cite{Lonsdale_2006} for a review). ULIRGs are typically interacting galaxies containing large amounts of heated gas and dust. They are mostly powered by starbursts with star-formation rates exceeding 100 $M_{\odot}~ \mathrm{yr^{-1}}$, with a possible secondary contribution from an active galactic nucleus (AGN). Such an AGN is likely Compton-thick, i.e.~it is obscured by matter with column densities $N_H \gtrsim 1.5 \times 10^{24}~ \mathrm{cm^{-2}}$ \cite{Comastri_2004}. Both starbursts and AGN are candidate environments for hadronic acceleration and neutrino production \cite{Tamborra_2014,Murase_2017}. In addition, ULIRGs are relatively abundant, with a local source density of $10^{-7}$--$10^{-6}~ \mathrm{Mpc^{-3}}$ which increases rapidly up to a redshift $z \sim 1$ \cite{Magnelli_2011}. As such, ULIRGs form a source class that could be responsible for a significant fraction of the diffuse astrophysical neutrino flux \cite{He_2013,Palladino_2019,Vereecken_2020}.

In this work we present the results of an IceCube stacking analysis searching for high-energy neutrinos from ULIRGs. These results will be extrapolated to the full ULIRG population, and they will also be compared to model predictions. For this we will use the \textit{Planck} 2015 results \cite{Planck_2015} for a $\Lambda$CDM cosmology, $H_0 = 67.8~ \mathrm{km~ s^{-1}~ Mpc^{-1}}$, $\Omega_{m} = 0.31$, and $\Omega_{\Lambda} = 0.69$. More information on the study presented in this work can be found in a recently submitted publication \cite{IceCube_ULIRG_paper}.

\section{ULIRG Stacking Analysis}

\subsection{Object Selection}

We select our ULIRGs from three catalogs that are primarily based on data from the Infrared Astronomical Satellite (\textit{IRAS}) \cite{Sanders_2003,Kim_1998,Nardini_2010}. The catalogs were already extensively described in previous proceedings \cite{ULIRG_analysis_ICRC2019}, as well the usage of the NASA/IPAC Extragalactic Database\footnote{The NASA/IPAC Extragalactic Database (NED) is operated by the Jet Propulsion Laboratory, California Institute of Technology, under contract with the National Aeronautics and Space Administration.} (NED) \cite{NED} to cross-identify these catalogs and obtain an initial selection of 189 ULIRGs. However, this initial selection is not fully representative for the local ULIRG population. In order to obtain such a representative sample, we make a completeness cut on our initial ULIRG selection, by only selecting those objects with a redshift $z \leq 0.13$. This redshift represents a conservative estimate of the distance up to which the least luminous ULIRG ($L_{\mathrm{IR}} = 10^{12} L_{\odot}$) can be observed given an \textit{IRAS} sensitivity $f_{60} = 1$ Jy, with $f_{60}$ the IR flux density at 60 $\mu$m. This implies that the redshift cut $z \leq 0.13$ essentially corresponds to a flux constraint $f_{60} \gtrsim 1$ Jy. The result is a sample of 75 ULIRGs.

Due to a limited sky coverage of the original ULIRG catalogs, which results from correlations with other surveys (see \cite{Kim_1998,Nardini_2010}), this redshift limited sample of 75 ULIRGs is not complete. It misses $\sim$40 sources with a flux in the range $1~\mathrm{Jy} < f_{60} < 5.24~\mathrm{Jy}$. Nevertheless, we estimate that the total stacking weight (see Section~\ref{sec:analysis_method}) of these $\sim$40 missing sources in our analysis is approximately 10\%. Therefore, our selection of 75 ULIRGs can still be regarded as a representative sample of the local ULIRG population.

\subsection{Detector and Data Set}

The $1~ \mathrm{km}^3$ IceCube Neutrino Observatory, located at the geographic South Pole, contains 5160 optical modules which detect the optical Cherenkov radiation emitted by secondary charged particles produced in the interactions of neutrinos\footnote{Since IceCube can in general not distinguish between neutrinos and antineutrinos, we use the term neutrino for both in these proceedings.} in the surrounding ice or the nearby bedrock. Using the detected Cherenkov light, the direction, energy and flavor of the neutrino can be reconstructed \cite{IceCube_detector}. In particular, muon neutrinos interacting via the charged-current interaction will produce secondary muons which leave track-like signatures in the detector, which yield an angular resolution $\lesssim 1^\circ$ for muon energies above 1 TeV \cite{IceCube_realtime_alert}.

The main backgrounds for astrophysical neutrino searches are atmospheric muons (at the kHz level) and atmospheric neutrinos (at the mHz level) produced in cosmic-ray air showers. In this work, we make use of the IceCube gamma-ray follow-up (GFU) data sample \cite{IceCube_realtime_alert}, which contains well-reconstructed muon tracks with a $4\pi$ sky coverage. The event selection of the GFU sample is performed separately for the Southern and Northern hemispheres. This is due to the fact that atmospheric muons cannot penetrate the Earth and only form a background in the Southern sky. Consequently, IceCube has a better sensitivity for astrophysical neutrinos originating from sources in the Northern sky. The GFU sample contains over 1.5 million events, spanning a livetime of 7.5 years of the full 86-string IceCube configuration.

\subsection{Analysis Method}
\label{sec:analysis_method}

In order to search for astrophysical neutrinos that are spatially correlated with our representative sample of $M=75$ ULIRGs, we perform a time-integrated unbinned maximum-likelihood analysis \cite{Braun_2008}. We also stack the contributions of the selected ULIRGs to enhance the sensitivity of the analysis \cite{IceCube_stacking}. The likelihood is given by
\begin{equation}
    \mathcal{L}(n_s,\gamma) = \prod_{i=1}^N \left[ \frac{n_s}{N} \sum_{k=1}^{M} w_k \mathcal{S}_i^k(\gamma) + \left( 1 - \frac{n_s}{N} \right) \mathcal{B}_i \right],
\label{eq:llh}
\end{equation}
where we fit for the number of astrophysical neutrinos, $n_s$, and the spectral index of such a signal component, $\gamma$, which is assumed to follow the same unbroken power-law spectrum for all ULIRGs. The factors $\mathcal{S}_i^k$ and $\mathcal{B}_i$ are the probability distribution functions (PDFs) corresponding with the signal of source $k$ and the background, respectively, evaluated for data event $i \in \lbrace 1,2,...,N \rbrace$. Both the energy and spatial components of the background PDF are constructed from experimental data. The signal PDF is determined from simulations, with a spatial component that is assumed to follow a two-dimensional Gaussian PDF centered around source $k$.

The contributions of all ULIRGs are stacked through the weighed sum over the signal PDFs of each source $k$. The stacking weight $w_k = r_k t_k / \sum_{j=1}^{M} r_j t_j$ depends on the detector response $r_k$ and a theoretical weight $t_k$. For the theoretical weight we use the total IR flux (8--1000 $\mu$m), $F_{\mathrm{IR}} = L_{\mathrm{IR}}/(4\pi d_L^2)$, where $d_L$ is the luminosity distance determined from the redshift measurements found in NED.

We then construct a test statistic as
\begin{equation}
    \mathrm{TS} = 2 \log \left[ \frac{\mathcal{L}(n_s=\hat{n}_s,\gamma=\hat{\gamma})}{\mathcal{L}(n_s=0)} \right],
\end{equation}
where $\hat{n}_s$ and $\hat{\gamma}$ are those values that maximize the likelihood. By construction, the TS provides a measure of how compatible the data is with a background-only scenario ($n_s = 0$), which is our null hypothesis. We construct the background-only PDF of the TS by scrambling data in right ascension $10^5$ times and computing the TS in each scramble. This PDF is well-described by a $\chi^2$ distribution, which is in accordance with Wilks' theorem \cite{Wilks_1938}.

The performance of the analysis can be tested by simulating pseudo-signal events according to an unbroken power-law spectrum, $\Phi_{\nu_\mu + \bar{\nu}_\mu}(E_\nu) = \Phi_{\nu_\mu + \bar{\nu}_\mu}(E_0) (E_\nu/E_0)^{-\gamma}$.
We define the sensitivity at 90\% confidence level (CL) as the number of pseudo-signal events required to obtain a p-value $\leq 0.5$ in 90\% of the pseudo-experiments. Analogously, we define the $3\sigma$ and $5\sigma$ discovery potentials as the number of pseudo-signal events required to obtain a p-value $\leq 2.70 \times 10^{-3}$ and $\leq 5.73 \times 10^{-7}$ in 50\% of the trials, respectively. Figure \ref{fig:sensitivity} shows these quantities as a function of the spectral index $\gamma$, both in terms of flux at the normalization energy $E_0 = 10$ TeV as well as in number of neutrinos. We note that our analysis is sensitive for neutrinos at $E_0 = 10$ TeV for all spectra considered in this work.

\section{Results and Interpretation}

\subsection{Analysis Results}

\begin{figure}[ht]
	\begin{center}
		\begin{subfigure}[b]{0.48\textwidth}
			\begin{center}
				\includegraphics[width=\textwidth]{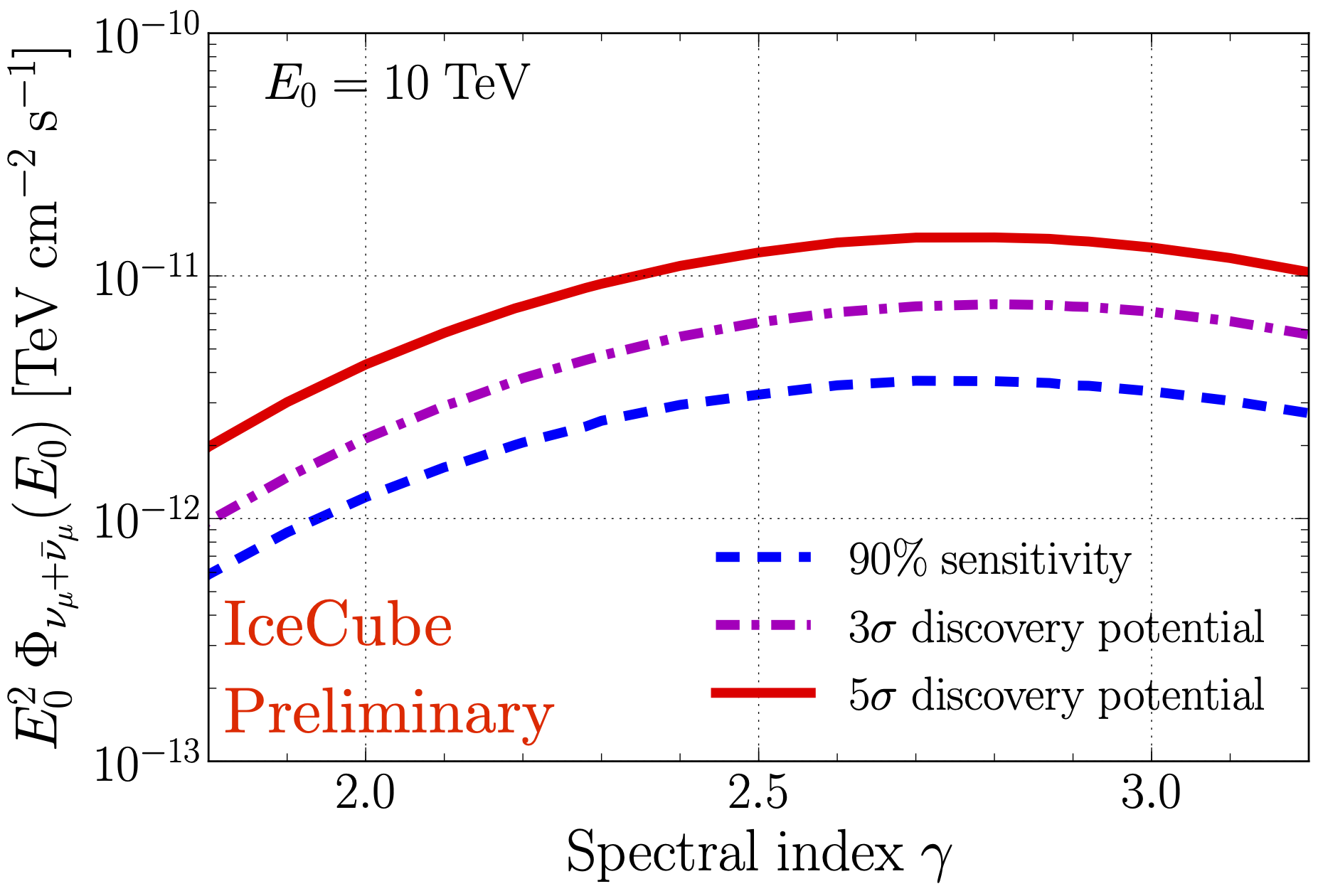}
			\end{center}
		\end{subfigure}
		\hspace{0.02\textwidth}
		\begin{subfigure}[b]{0.48\textwidth}
			\begin{center}
				\includegraphics[width=\textwidth]{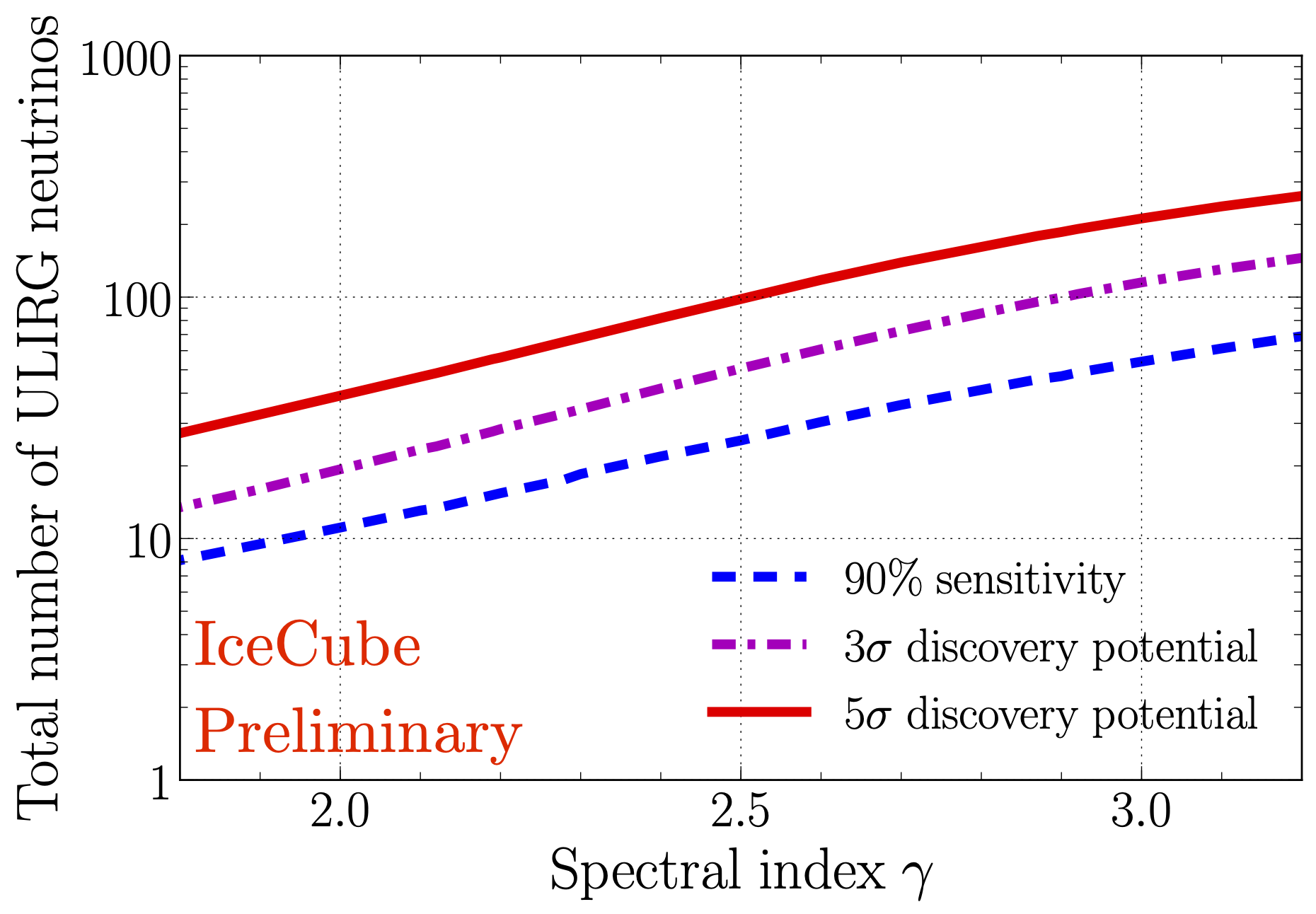}
			\end{center}
		\end{subfigure}
		\caption{\textit{Left}: Sensitivity, $3\sigma$, and $5\sigma$ discovery potentials as a function of the spectral index $\gamma$, for an unbroken $E^{-\gamma}$ power-law spectrum, in terms of the stacked flux at the normalization energy $E_0 = 10~ \mathrm{TeV}$. \textit{Right}: Same as left plot but in terms of the total number of ULIRG neutrinos.}
		\label{fig:sensitivity}
	\end{center}
\end{figure}

The stacking analysis yields a best-fit for the number of signal neutrinos $\hat{n}_s = 0$, which leaves the best-fit spectral index $\hat{\gamma}$ undetermined. This also corresponds with a $\mathrm{TS} = 0$ and a p-value = 1. Hence, our results are compatible with the background-only hypothesis. Consequently, we compute upper limits at 90\% confidence level (CL) on the stacked muon-neutrino flux originating from our selection of 75 ULIRGs. These upper limits coincide with the sensitivities of the analysis, shown in Figure \ref{fig:sensitivity} at $E_0 = 10$ TeV.

\subsection{Limits on the ULIRG Source Population}

The stacked flux upper limits can be converted to limits on the diffuse neutrino flux from the population of ULIRGs with $z \leq 0.13$ as $\Phi_{\nu_{\mu} + \bar{\nu}_{\mu}}^{z\leq0.13} = \epsilon_c \Phi_{\nu_{\mu} + \bar{\nu}_{\mu}}^{90\%} / (4\pi~ \mathrm{sr})$. The factor $\epsilon_c = 1.1$ takes into account the contribution from sources with $z \leq 0.13$ that are missed due to the limited sky coverage of the ULIRG catalogs used for our object selection. These limits can then be extrapolated to the diffuse flux of the ULIRG population with $z \leq z_{\max}$,
\begin{equation}
    \Phi_{\nu_{\mu} + \bar{\nu}_{\mu}}^{z \leq z_{\max}} = \frac{\xi_{z=z_{\max}}}{\xi_{z=0.13}} \Phi_{\nu_{\mu} + \bar{\nu}_{\mu}}^{z \leq 0.13}.
\end{equation}
Here we assume that over cosmic history, all ULIRGs have identical properties of hadronic acceleration and neutrino production. The factor $\xi_z$, which becomes energy-independent for the power-law spectra considered in this work, takes into account the redshift evolution of the sources \cite{Ahlers_2014}. We parameterize the ULIRG redshift evolution according to \cite{Vereecken_2020}, $\mathcal{H}(z) = (1+z)^m$, with $m=4$ for $0 \leq z < 1$ and $m=0$ for $z \geq 1$.

\begin{figure}[ht]
	\begin{center}
		\begin{subfigure}[b]{0.48\textwidth}
			\begin{center}
				\includegraphics[width=\textwidth]{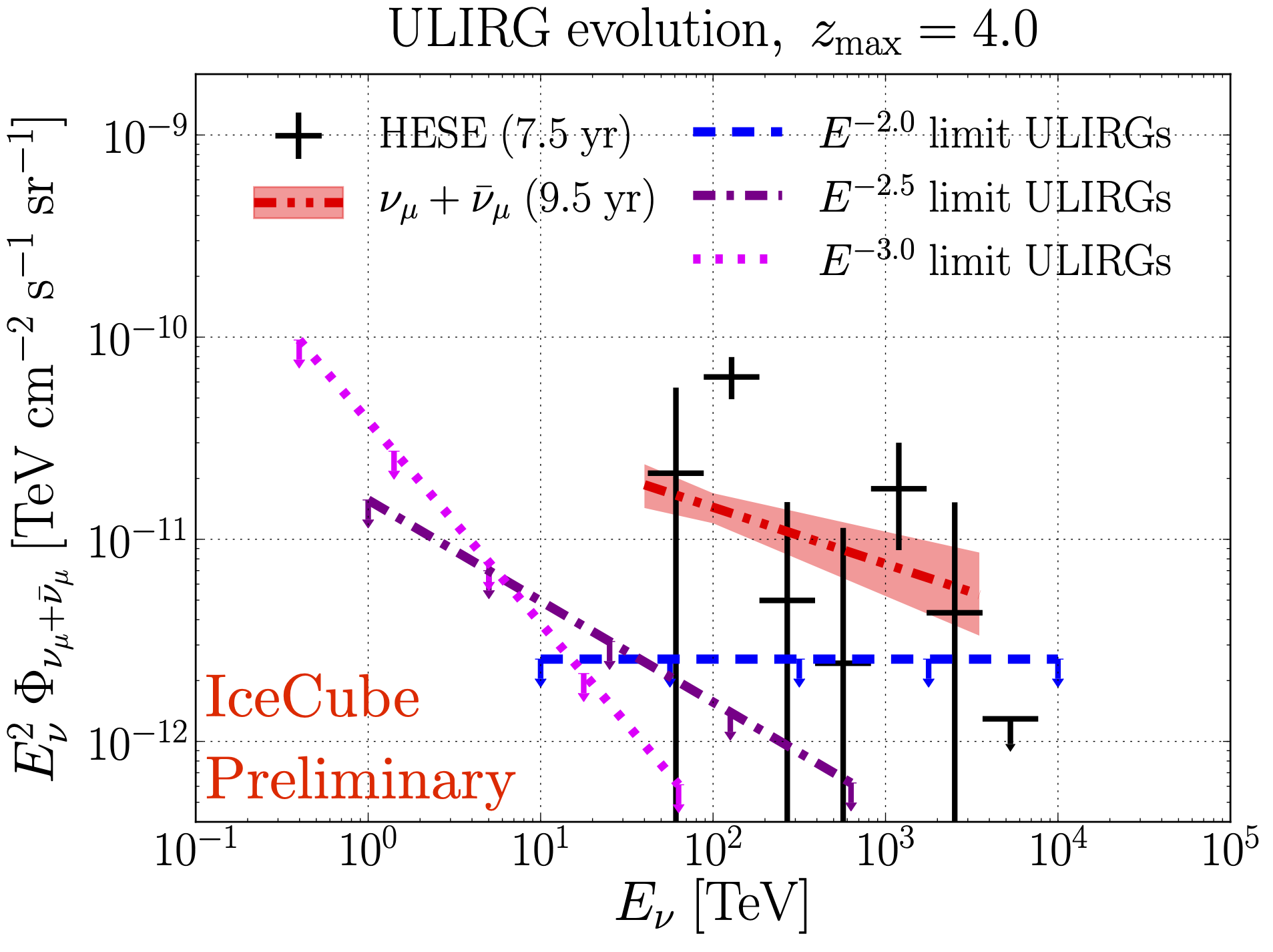}
			\end{center}
		\end{subfigure}
		\hspace{0.02\textwidth}
		\begin{subfigure}[b]{0.48\textwidth}
			\begin{center}
				\includegraphics[width=\textwidth]{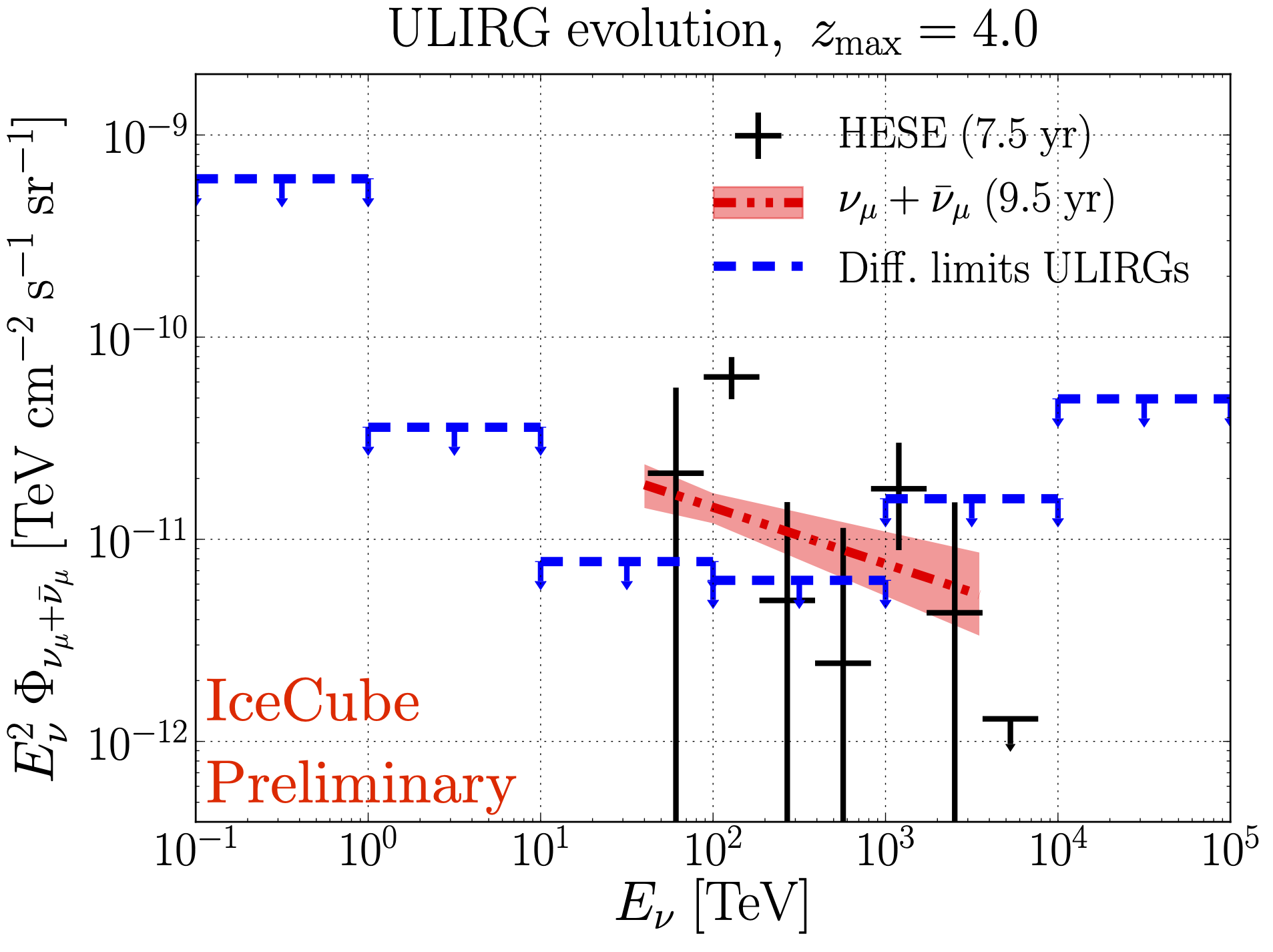}
			\end{center}
		\end{subfigure}
		\caption{\textit{Left}: Integral limits at 90\% confidence level on the contribution of the ULIRG population up to a redshift $z_{\max} = 4.0$ to the observed diffuse neutrino flux of \cite{IceCube_diffuse_numu_9.5yr,IceCube_diffuse_HESE_7.5yr}. These are shown for unbroken $E^{-2.0}$ (dashed blue line), $E^{-2.5}$ (dash-dotted dark magenta line), and $E^{-3.0}$ (dotted light magenta line) power-law spectra. The integral limits are plotted within their respective 90\% central energy ranges. \textit{Right}: Quasi-differential limits at 90\% confidence level (dashed blue line), found by determining the limit for an $E^{-2.0}$ spectrum in each decade of energy.}
		\label{fig:population_limits}
	\end{center}
\end{figure}

In Figure \ref{fig:population_limits} we report the upper limits at 90\% CL on the population of ULIRGs up to a redshift $z_{\max} = 4.0$ following \cite{Palladino_2019}. The left panel of Figure \ref{fig:population_limits} shows these in terms of the integral limits for unbroken $E^{-2.0}$, $E^{-2.5}$, and $E^{-3.0}$ power-law spectra. These are shown within the respective 90\% central energy ranges that contribute to the upper limits. We find that the $E^{-2.0}$ and $E^{-2.5}$ constrain the ULIRG contribution to the diffuse neutrino observations of \cite{IceCube_diffuse_numu_9.5yr,IceCube_diffuse_HESE_7.5yr} up to $\sim$3 PeV and $\sim$600 TeV, respectively. The right panel of Figure \ref{fig:population_limits} shows the quasi-differential limits, computed by determining the $E^{-2.0}$ limit per bin of energy decade. Also here we find that our stacking analysis constrains the contribution of ULIRGs to the diffuse observations for the energy bins 10--100 TeV and 0.1--1 PeV. Note that this interpretation does not take into account a possible contribution from the less luminous but more numerous Luminous Infrared Galaxies (LIRGs, $L_{\mathrm{IR}} \geq 10^{11} L_{\odot}$).

\subsection{Comparison with Model Predictions}

\begin{figure}[ht]
	\begin{center}
		\begin{subfigure}[b]{0.48\textwidth}
			\begin{center}
				\includegraphics[width=\textwidth]{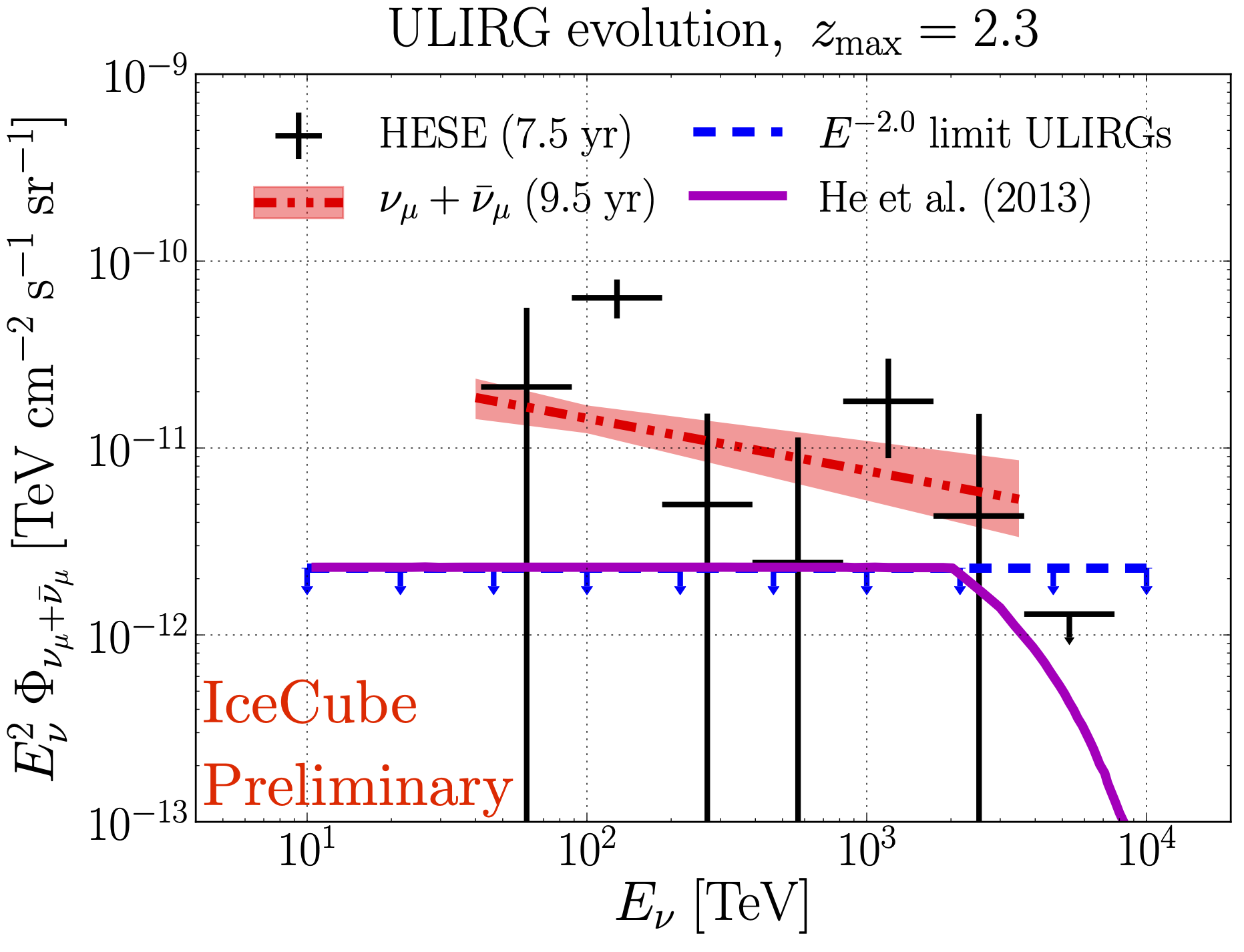}
			\end{center}
		\end{subfigure}
		\hspace{0.02\textwidth}
		\begin{subfigure}[b]{0.48\textwidth}
			\begin{center}
				\includegraphics[width=\textwidth]{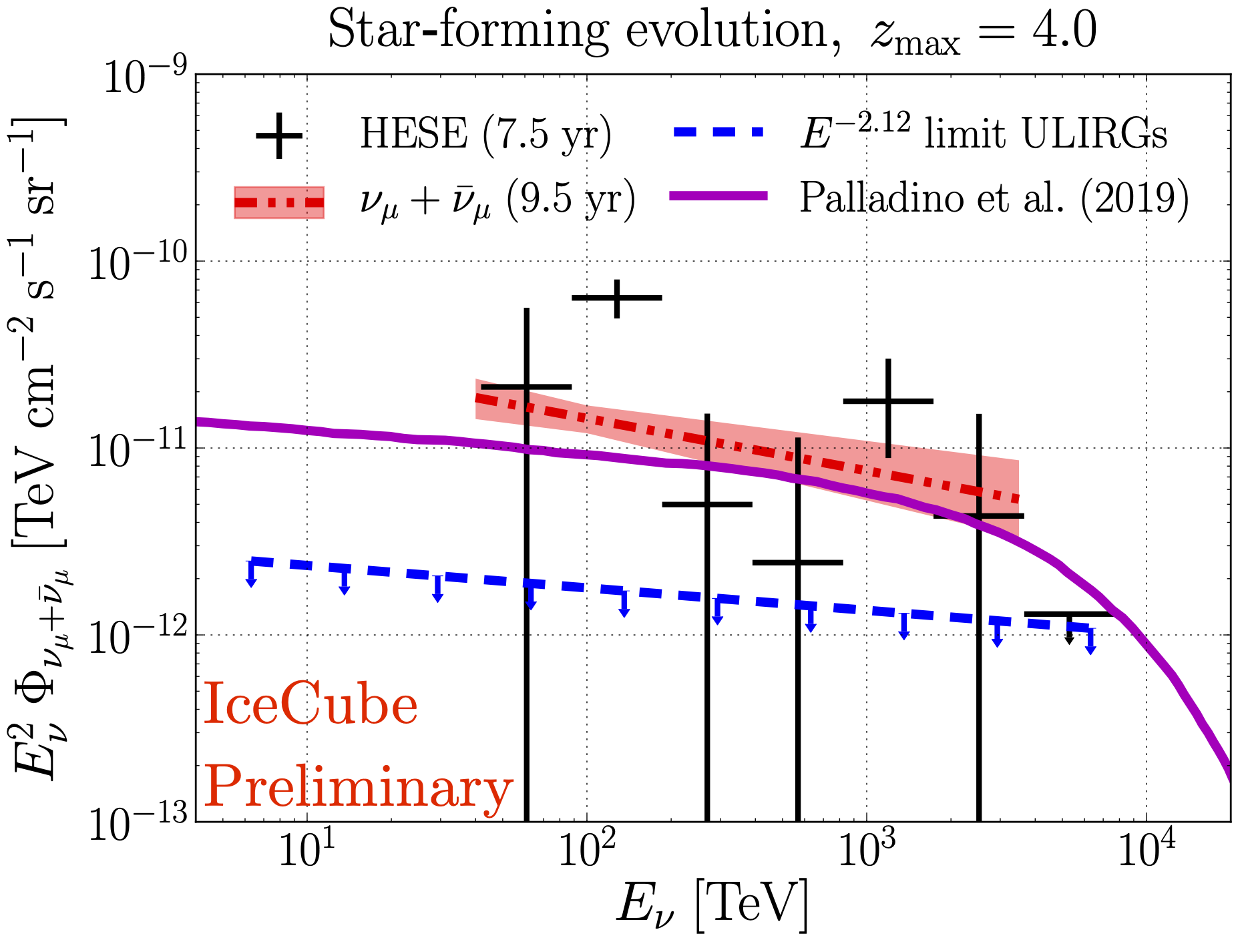}
			\end{center}
		\end{subfigure}
		\caption{\textit{Left}: Prediction by \cite{He_2013} of a diffuse neutrino flux from hypernovae in ULIRGs (full magenta line) and the $E^{-2.0}$ upper limit at 90\% confidence level of this search (dashed blue line). For comparison, an ULIRG source evolution is assumed and integrated up to a redshift $z_{\max} = 2.3$. \textit{Right}: Prediction by \cite{Palladino_2019} of a diffuse neutrino flux, normalized to the IceCube observations, from hadronically powered gamma-ray galaxies (HAGS; full magenta line) and the $E^{-2.12}$ upper limit at 90\% confidence level of this search (dashed blue line). For comparison, a source evolution according to the star-formation rate is assumed and integrated up to a redshift $z_{\max} = 4.0$.}
		\label{fig:model_limits}
	\end{center}
\end{figure}

We first compare our results with the starburst reservoir model of He et al.~\cite{He_2013}, who consider hypernovae as hadronic accelerators. They predict a diffuse neutrino flux from ULIRGs up to $z_{\max} = 2.3$, which we compare to our $E^{-2.0}$ upper limit in the left panel of Figure \ref{fig:model_limits}. We find that the He et al.~prediction is at the level of our upper limit; more years of data are required in order to validate or constrain this model.

The second reservoir model we consider is that of Palladino et al.~\cite{Palladino_2019}. They construct a generic framework to compute diffuse neutrino and gamma-ray fluxes from a population of hadronically-powered gamma-ray galaxies (HAGS), such as ULIRGs or starburst galaxies with $L_{\mathrm{IR}} < 10^{12} L_{\odot}$. They find that for power-law spectra with an index $\gamma \leq 2.12$ and an exponential cutoff at $\sim$10 PeV, HAGS can fit the bulk of the diffuse neutrino observations whilst remaining consistent with the non-blazar EGB bound above 50 GeV \cite{Fermi_EGB}. We compare their most optimistic scenario of $\gamma = 2.12$ with our $E^{-2.12}$ upper limit in the right panel of Figure \ref{fig:model_limits}. Here we follow Palladino et al.~and take $z_{\max} = 4.0$ and the redshift evolution, according to the star-formation rate \cite{Yuksel_2008}, $\mathcal{H}(z) = (1+z)^m$ with $m=3.4$ for $0 \leq z < 1$ and $m=-0.3$ for $1 < z \leq 4$. We find that ULIRGs cannot be the sole population of HAGS that contributes to the diffuse neutrino observations. Note that this does not have any implications for other candidate HAGS populations.

Lastly, we make a comparison with the AGN beam-dump model of Vereecken \& de Vries \cite{Vereecken_2020}. They consider a possible Compton-thick AGN in ULIRGs as the hadronic accelerator, where the neutrino production occurs in the interactions of accelerated hadrons with the obscuring dust and gas. They find that ULIRGs can fit the diffuse IceCube observations without exceeding the non-blazar EGB bound above 50 GeV \cite{Fermi_EGB} if the obscuring matter has a column density $N_H \gtrsim 5\times 10^{25}~ \mathrm{cm^{-2}}$, since then the gamma rays are attenuated before escaping the source. However, this model strongly depends on the electron-to-proton luminosity ratio $f_e$ \cite{Merten_2017}, which remains a highly uncertain parameter. By fitting the prediction of Vereecken \& de Vries to our $E^{-2.0}$ upper limit on the ULIRG neutrino flux up to $z_{\max} = 4.0$, we can roughly estimate the lower limit $f_e \gtrsim 10^{-3}$. However, due to other uncertainties in the model, this lower limit should be regarded as an order-of-magnitude estimation. Nevertheless, we remark that this value is consistent with previous limits on $f_e$ provided in \cite{Vereecken_2020} in the context of an obscured-AGN analysis performed with IceCube \cite{Maggi_2016,IceCube_obscured_AGN}.

\section{Conclusions}

We presented the results of an IceCube stacking analysis searching for neutrinos from a representative sample of 75 ULIRGs within a redshift $z \leq 0.13$. No astrophysical component was found in the data, which was consistent with background. We therefore set upper limits on the stacked muon-neutrino flux originating from our representative ULIRG sample. These limits were extrapolated to a diffuse neutrino limit on the full ULIRG population. We found that the contribution of ULIRGs to the observed diffuse neutrino flux is constrained up to $\sim$3 PeV and $\sim$600 TeV for unbroken $E^{-2.0}$ and $E^{-2.5}$ power-law spectra, respectively. Also the quasi-differential limits provide stronger constraints than the diffuse flux in the energy ranges 10--100 TeV and 0.1--1 PeV. We conclude that the population of ULIRGs cannot be the sole contributors to the diffuse neutrino observations.

In addition, we compared our results with three model predictions. First, we found that the prediction by He et al.~\cite{He_2013} is at the level of our $E^{-2.0}$ upper limit. Second, we excluded ULIRGs as the sole HAGS that could be responsible for the diffuse neutrino observations in the context of the model by Palladino et al.~\cite{Palladino_2019}. Last, we considered the AGN beam-dump model by Vereecken \& de Vries \cite{Vereecken_2020}, which mainly suffers from uncertainties through the electron-to-proton luminosity ratio $f_e$. As such, we made an order-of-magnitude estimation on the lower limit of this parameter, $f_e \gtrsim 10^{-3}$.

\small
\bibliographystyle{ICRC}
\bibliography{references}

\providecommand{\href}[2]{#2}\begingroup\raggedright\begin{thebibliography}{10}

\bibitem{IceCube_discovery_science}
{\bfseries IceCube} Collaboration, M.~Aartsen {\em et~al.}
  \href{http://dx.doi.org/10.1126/science.1242856}{{\em Science} {\bfseries
  342} (2013) 1242856}.

\bibitem{Murase_2016}
K.~Murase and E.~Waxman
  \href{http://dx.doi.org/10.1103/PhysRevD.94.103006}{{\em Phys.~Rev.~D}
  {\bfseries 94} (2016) 103006}.

\bibitem{Fermi_LAT}
{\bfseries \textit{Fermi}-LAT} Collaboration, W.~Atwood {\em et~al.}
  \href{http://dx.doi.org/10.1088/0004-637X/697/2/1071}{{\em ApJ} {\bfseries
  697} (2009) 1071--1102}.

\bibitem{IceCube_2LAC_constraints}
{\bfseries IceCube} Collaboration, M.~Aartsen {\em et~al.}
  \href{http://dx.doi.org/10.3847/1538-4357/835/1/45}{{\em ApJ} {\bfseries 835}
  no.~1, (2017) 45}.

\bibitem{Fermi_EGB}
{\bfseries \textit{Fermi}-LAT} Collaboration, M.~Ackermann {\em et~al.}
  \href{http://dx.doi.org/10.1103/PhysRevLett.116.151105}{{\em
  Phys.~Rev.~Lett.} {\bfseries 116} no.~15, (2016) 151105}.

\bibitem{Bechtol_2017}
K.~Bechtol {\em et~al.}
  \href{http://dx.doi.org/10.3847/1538-4357/836/1/47}{{\em ApJ} {\bfseries 836}
  no.~1, (2017) 47}.

\bibitem{Murase_2016_hidden_sources}
K.~Murase, D.~Guetta, and M.~Ahlers
  \href{http://dx.doi.org/10.1103/PhysRevLett.116.071101}{{\em
  Phys.~Rev.~Lett.} {\bfseries 116} no.~7, (2016) 071101}.

\bibitem{Lonsdale_2006}
C.~J. Lonsdale, D.~Farrah, and H.~E. Smith, {\em Ultraluminous Infrared
  Galaxies}, \href{http://dx.doi.org/10.1007/3-540-30313-8_9}{pp.~285--336}.
\newblock Springer Berlin Heidelberg, Berlin, Heidelberg, 2006.

\bibitem{Comastri_2004}
A.~Comastri \href{http://dx.doi.org/10.1007/978-1-4020-2471-9_8}{{\em ASSL}
  {\bfseries 308} (2004) 245}.

\bibitem{Tamborra_2014}
I.~Tamborra, S.~Ando, and K.~Murase
  \href{http://dx.doi.org/10.1088/1475-7516/2014/09/043}{{\em JCAP} {\bfseries
  09} (2014) 043}.

\bibitem{Murase_2017}
K.~Murase, {\em {Active Galactic Nuclei as High-Energy Neutrino Sources}},
  \href{http://dx.doi.org/10.1142/9789814759410_0002}{pp.~15--31}.
\newblock World Scientific, 2017.

\bibitem{Magnelli_2011}
B.~Magnelli {\em et~al.}
  \href{http://dx.doi.org/10.1051/0004-6361/200913941}{{\em A\&A} {\bfseries
  528} (2011) A35}.

\bibitem{He_2013}
H.-N. He {\em et~al.} \href{http://dx.doi.org/10.1103/PhysRevD.87.063011}{{\em
  Phys.~Rev.~D} {\bfseries 87} no.~6, (2013) 063011}.

\bibitem{Palladino_2019}
A.~Palladino {\em et~al.}
  \href{http://dx.doi.org/10.1088/1475-7516/2019/09/004}{{\em JCAP} {\bfseries
  09} (2019) 004}.

\bibitem{Vereecken_2020}
M.~Vereecken and K.~D. de~Vries
  \href{http://arxiv.org/abs/2004.03435}{{\ttfamily arXiv:2004.03435
  [astro-ph.HE]}}.

\bibitem{Planck_2015}
{\bfseries \textit{Planck}} Collaboration, P.~Ade {\em et~al.}
  \href{http://dx.doi.org/10.1051/0004-6361/201525830}{{\em A\&A} {\bfseries
  594} (2016) A13}.

\bibitem{IceCube_ULIRG_paper}
{\bfseries IceCube} Collaboration, R.~Abbasi {\em et~al.}
  \href{http://arxiv.org/abs/2107.03149}{{\ttfamily arXiv:2107.03149
  [astro-ph.HE]}}.

\bibitem{Sanders_2003}
D.~Sanders {\em et~al.} \href{http://dx.doi.org/10.1086/376841}{{\em AJ}
  {\bfseries 126} (2003) 1607}.

\bibitem{Kim_1998}
D.-C. Kim and D.~Sanders \href{http://dx.doi.org/10.1086/313148}{{\em ApJS}
  {\bfseries 119} (1998) 41--58}.

\bibitem{Nardini_2010}
E.~Nardini {\em et~al.}
  \href{http://dx.doi.org/10.1111/j.1365-2966.2010.16618.x}{{\em MNRAS}
  {\bfseries 405} (2010) 2505}.

\bibitem{ULIRG_analysis_ICRC2019}
{\bfseries IceCube} Collaboration, P.~Correa, K.~D. de~Vries, and N.~van
  Eijndhoven \href{http://dx.doi.org/10.22323/1.358.0860}{{\em PoS} {\bfseries
  ICRC2019} (2020) 860}.

\bibitem{NED}
{NASA/IPAC Extragalactic Database, \href{http://ned.ipac.caltech.edu}{\tt
  http://ned.ipac.caltech.edu}}.

\bibitem{IceCube_detector}
{\bfseries IceCube} Collaboration, M.~Aartsen {\em et~al.}
  \href{http://dx.doi.org/10.1088/1748-0221/12/03/P03012}{{\em JINST}
  {\bfseries 12} no.~03, (2017) P03012}.

\bibitem{IceCube_realtime_alert}
{\bfseries IceCube} Collaboration, M.~Aartsen {\em et~al.}
  \href{http://dx.doi.org/10.1016/j.astropartphys.2017.05.002}{{\em
  Astropart.~Phys.} {\bfseries 92} (2017) 30--41}.

\bibitem{Braun_2008}
J.~Braun {\em et~al.}
  \href{http://dx.doi.org/10.1016/j.astropartphys.2008.02.007}{{\em
  Astropart.~Phys.} {\bfseries 29} (2008) 299--305}.

\bibitem{IceCube_stacking}
{\bfseries IceCube} Collaboration, A.~Achterberg {\em et~al.}
  \href{http://dx.doi.org/10.1016/j.astropartphys.2006.06.012}{{\em
  Astropart.~Phys.} {\bfseries 26} (2006) 282--300}.

\bibitem{Wilks_1938}
S.~Wilks \href{http://dx.doi.org/10.1214/aoms/1177732360}{{\em Annals
  Math.~Statist.} {\bfseries 9} no.~1, (1938) 60--62}.

\bibitem{Ahlers_2014}
M.~Ahlers and F.~Halzen
  \href{http://dx.doi.org/10.1103/PhysRevD.90.043005}{{\em Phys.~Rev.~D}
  {\bfseries 90} no.~4, (2014) 043005}.

\bibitem{IceCube_diffuse_numu_9.5yr}
{\bfseries IceCube} Collaboration, J.~Stettner
  \href{http://dx.doi.org/10.22323/1.358.1017}{{\em PoS} {\bfseries ICRC2019}
  (2020) 1017}.

\bibitem{IceCube_diffuse_HESE_7.5yr}
{\bfseries IceCube} Collaboration, R.~Abbasi {\em et~al.}
  \href{http://arxiv.org/abs/2011.03545}{{\ttfamily arXiv:2011.03545
  [astro-ph.HE]}}.

\bibitem{Yuksel_2008}
H.~Y{\"{u}}ksel {\em et~al.} \href{http://dx.doi.org/10.1086/591449}{{\em ApJL}
  {\bfseries 683} (2008) L5--L8}.

\bibitem{Merten_2017}
L.~Merten {\em et~al.}
  \href{http://dx.doi.org/10.1016/j.astropartphys.2017.02.007}{{\em
  Astropart.~Phys.} {\bfseries 90} (2017) 75--84}.

\bibitem{Maggi_2016}
G.~Maggi {\em et~al.} \href{http://dx.doi.org/10.1103/PhysRevD.94.103007}{{\em
  Phys.~Rev.~D} {\bfseries 94} no.~10, (2016) 103007}.

\bibitem{IceCube_obscured_AGN}
{\bfseries IceCube} Collaboration, G.~Maggi, K.~D. de~Vries, and N.~van
  Eijndhoven \href{http://dx.doi.org/10.22323/1.301.1000}{{\em PoS} {\bfseries
  ICRC2017} (2018) 1000}.

\end{thebibliography}\endgroup

\clearpage
\section*{Full Author List: IceCube Collaboration}




\scriptsize
\noindent
R. Abbasi$^{17}$,
M. Ackermann$^{59}$,
J. Adams$^{18}$,
J. A. Aguilar$^{12}$,
M. Ahlers$^{22}$,
M. Ahrens$^{50}$,
C. Alispach$^{28}$,
A. A. Alves Jr.$^{31}$,
N. M. Amin$^{42}$,
R. An$^{14}$
% ...
\\

\noindent
% ...
$^{12}$ Universit{\'e} Libre de Bruxelles, Science Faculty CP230, B-1050 Brussels, Belgium \\
% ...
$^{14}$ Department of Physics and Laboratory for Particle Physics and Cosmology, Harvard University, Cambridge, MA 02138, USA \\
% ...
$^{17}$ Department of Physics, Loyola University Chicago, Chicago, IL 60660, USA \\
$^{18}$ Dept. of Physics and Astronomy, University of Canterbury, Private Bag 4800, Christchurch, New Zealand \\
% ...
$^{22}$ Niels Bohr Institute, University of Copenhagen, DK-2100 Copenhagen, Denmark \\
% ...
$^{28}$ D{\'e}partement de physique nucl{\'e}aire et corpusculaire, Universit{\'e} de Gen{\`e}ve, CH-1211 Gen{\`e}ve, Switzerland \\
% ...
$^{31}$ Karlsruhe Institute of Technology, Institute for Astroparticle Physics, D-76021 Karlsruhe, Germany  \\
% ...
$^{42}$ Bartol Research Institute and Dept. of Physics and Astronomy, University of Delaware, Newark, DE 19716, USA \\
% ...
$^{50}$ Oskar Klein Centre and Dept. of Physics, Stockholm University, SE-10691 Stockholm, Sweden \\
% ...
$^{59}$ DESY, D-15738 Zeuthen, Germany \\
% ...

\subsection*{Acknowledgements}

\noindent
USA {\textendash} U.S. National Science Foundation-Office of Polar Programs,
U.S. National Science Foundation-Physics Division,
U.S. National Science Foundation-EPSCoR,
Wisconsin Alumni Research Foundation,
Center for High Throughput Computing (CHTC) at the University of Wisconsin{\textendash}Madison,
Open Science Grid (OSG),
Extreme Science and Engineering Discovery Environment (XSEDE),
Frontera computing project at the Texas Advanced Computing Center,
U.S. Department of Energy-National Energy Research Scientific Computing Center,
Particle astrophysics research computing center at the University of Maryland,
Institute for Cyber-Enabled Research at Michigan State University,
and Astroparticle physics computational facility at Marquette University;
Belgium {\textendash} Funds for Scientific Research (FRS-FNRS and FWO),
FWO Odysseus and Big Science programmes,
and Belgian Federal Science Policy Office (Belspo);
Germany {\textendash} Bundesministerium f{\"u}r Bildung und Forschung (BMBF),
Deutsche Forschungsgemeinschaft (DFG),
Helmholtz Alliance for Astroparticle Physics (HAP),
Initiative and Networking Fund of the Helmholtz Association,
Deutsches Elektronen Synchrotron (DESY),
and High Performance Computing cluster of the RWTH Aachen;
Sweden {\textendash} Swedish Research Council,
Swedish Polar Research Secretariat,
Swedish National Infrastructure for Computing (SNIC),
and Knut and Alice Wallenberg Foundation;
Australia {\textendash} Australian Research Council;
Canada {\textendash} Natural Sciences and Engineering Research Council of Canada,
Calcul Qu{\'e}bec, Compute Ontario, Canada Foundation for Innovation, WestGrid, and Compute Canada;
Denmark {\textendash} Villum Fonden and Carlsberg Foundation;
New Zealand {\textendash} Marsden Fund;
Japan {\textendash} Japan Society for Promotion of Science (JSPS)
and Institute for Global Prominent Research (IGPR) of Chiba University;
Korea {\textendash} National Research Foundation of Korea (NRF);
Switzerland {\textendash} Swiss National Science Foundation (SNSF);
United Kingdom {\textendash} Department of Physics, University of Oxford.

\end{document}